\documentclass[a4paper,11pt]{article}
\usepackage{pos}

\newcommand{\epsal}{\varepsilon_\alpha}

\newcommand{\di}{\text{d}}

\renewcommand{\logo}{\relax}

\title{From vacuum decay to gravitational waves}

\author*[a,b]{Marco Matteini}

\affiliation[a]{Jo\v{z}ef Stefan Institute,\\
Jamova 39, 1000 Ljubljana, Slovenia}

\affiliation[b]{Faculty of Mathematics and Physics, University of Ljubljana,\\
Jadranska 19, 1000 Ljubljana, Slovenia}

\emailAdd{marco.matteini@ijs.si}

\abstract{I present a recap of a fully analytical calculation of the Euclidean action for a self-interacting scalar field with a quartic potential, in the thin-wall approximation. I then apply this result to the coupled fluid-scalar field model, a phenomenologically relevant model for cosmological first order phase transitions, and compare results with numerical simulations from the literature.}

\FullConference{2nd Training School and General Meeting of the COST Action COSMIC WISPers  (CA21106) (COSMICWISPers2024)\\
 10-14 June 2024 and 3-6 September 2024\\
Ljubljana (Slovenia) and Istanbul (Turkey)\\}

\begin{document}

\maketitle

\section{Introduction} \label{introduction}
The detection of gravitational waves has opened up a new window to study our Universe, complementary to observations via electromagnetic radiation.
The majority of detected signals come from astrophysical events, such as mergers of black holes in binary systems, but cosmological events can also be a compelling source of gravitational waves.
Such an example is provided by first order phase transitions~\cite{Witten:1984rs}, where a stable phase at high temperatures becomes metastable (false vacuum) as the Universe cools down, opening up the possibility for a decay to the new global minimum (true vacuum) via tunneling.
These events are particularly interesting for two main reasons.
The first one is that they are directly linked to Beyond the Standard Model physics.
This is due to the fact that, within the Standard Model, the electroweak phase transition and the QCD phase transition are crossovers~\cite{Stephanov:2006zvm,Kajantie:1995kf}.
The second reason is that a number of gravitational wave experiments, such as LISA~\cite{LISA:2017pwj}, are on the horizon, and might be able to detect signals coming form such events~\cite{Caprini:2019egz}\footnote{In particular, LISA will be sensitive to frequencies in the mHz region, which corresponds to a transition happening around the electroweak scale.}.

A first order phase transition proceeds via bubble nucleation: spherical regions of spacetime decay from the false vacuum to the true vacuum configuration.
If such bubbles are large enough, it is energetically favourable for them to grow, so they will expand, merge with other bubbles, and convert the whole spacetime to the true vacuum configuration.
The physical quantity to describe bubble nucleation in a thermal field theory is the decay rate~\cite{Linde:1981zj}
\begin{equation} 
\Gamma(T) = T^4\left(\frac{S_3}{2\pi T}\right)^{3/2} \text{e}^{-S_3/T} \, ,
\end{equation}
where $S_3$ is the Euclidean action of the tunneling solution.

As the Universe keeps expanding and cooling down, more bubbles nucleate.
A relevant temperature that indicates the onset of the transition is the nucleation temperature $T_N$, defined as the temperature at which, on average, one bubble nucleates per Hubble volume
\begin{align} \label{eq:Tn}
    \int_{T_N}^{T_c}\frac{\di T}{T} \frac{\Gamma(T)}{H(T)^4} = 1 \, ,
\end{align}
where $T_c$ is the critical temperature, at which the minima are degenerate, and $H$ is the Hubble parameter.
A successful completion of the phase transition is indicated by the percolation temperature $T_p$, where a connected region of the true vacuum phase is present in the Universe.
This corresponds to the true vacuum volume being at least 34\% of the comoving volume
\begin{align}
    I(T_p) = \frac{4\pi v_w}{3}\int^{T_c}_{T_p} \di T' \,\frac{\Gamma(T')}{H(T')\,T'^4}\left(\int_{T_p}^{T'}\frac{\di T''}{H(T'')}\right)^3 \, =0.34 \, ,
\end{align}
where $v_w$ is the bubble wall velocity.

The transition is furthermore characterized by two parameters, which affect the amplitude and frequency of the gravitational wave power spectrum:
the inverse duration $\beta$, which is the time derivative of the decay rate at the percolation temperature,
and the strength $\alpha$ given by\footnote{We adopt here the definition used in~\cite{Hindmarsh:2015qta}, for the sake of comparison between the results obtained there and the ones presented here.}
\begin{equation} \label{eq:alpha}
\alpha=\frac{w(\phi_s,T)-w(\phi_b,T)}{3aT^4}\, ,
\end{equation}
where $\phi_s$ ($\phi_b$) indicates the symmetric (broken) phase, $a=g_*\pi^2/90$, $g_*$ is the number of relativistic degrees of freedom, and
\begin{equation}
\epsilon(\phi,T) = 3aT^4+V(\phi,T)-T\dfrac{\partial V}{\partial T} \, , \qquad p = aT^4-V(\phi,T) \, , \qquad w=\epsilon+p \, .
\end{equation}

This paper is organized as follows.
In Section \ref{sec:TW} I present a fully analytical calculation of the Euclidean action for a single scalar field in the thin-wall approximation, taken from~\cite{Matteini:2024xvg}.
In Section \ref{sec:FF} I apply this analytical result to a phenomenologically relevant scenario for a cosmological phase transition.
In Section \ref{conclusion} I summarize the results and conclusion.

\section{Thin and thick wall} \label{sec:TW}
In this Section I will follow~\cite{Matteini:2024xvg} and consider a phase transition driven by a single scalar with a quartic potential
\begin{equation}
\label{eq:VTW}
V_{TW}(\phi)=\frac{1}{2}m^2\phi^2+\eta\phi^3+\frac{1}{8}\lambda_{C}\phi^4 \, ,
\end{equation}
which features two minima, one at the origin and one away from the origin.
The parameters are chosen such that $m^2>0$, $\eta>0$ and $0<\lambda_C<4\eta^2/m^2$, so that the field value at the true vacuum is negative.
When $m^2$ approaches zero, the false vacuum and the maximum of the potential barrier merge into an inflection point, and there is only one minimum, the true vacuum\footnote{This limit is relevant for cosmological phase transitions that exhibit supercooling: such transitions take place at a temperature much lower than the critical temperature.}.

In Euclidean dimensions $D$, the tunneling profile enjoys an $O(D)$ symmetry.
The Euclidean action is given by
\begin{align}
S &= \Omega \int_0^\infty {\rm d}\rho \ \rho^{D-1} \left( \frac{1}{2} \left( 
  \frac{{\rm d}\phi}{{\rm d}\rho} \right)^2 + V(\phi) - V_{\rm FV} \right) \, ,
  &
  \Omega &= \frac{2 \pi^{D/2}}{\Gamma \left(D/2 \right)} \, ,
\end{align}
where $\Omega$ is the solid angle in $D$ dimensions, $\rho$ is the Euclidean radius and $V_\text{FV}$ denotes the value of the potential at the false vacuum.
The action is extremized to obtain the equations of motion
\begin{equation}
\ddot{\phi}+ \frac{D-1}{\rho} \dot \phi = \frac{{\rm d}V}{{\rm d}\phi} \, , \qquad \phi \left(\rho = \infty \right) = \phi_{\rm FV} \, , \qquad \dot \phi \left(\rho = 0 \right) = 0 \, ,
\end{equation}
where the dot denotes a derivative with respect to $\rho$.
For computational purposes, it is convenient to introduce dimensionless variables
\begin{align} 
  \varphi  \equiv \frac{2\eta}{m^2} \phi \, , 
  \qquad 
  \tilde\rho \equiv m \rho \, ,
  \qquad
  \epsal \equiv 1 - \lambda_C \frac{m^2}{4\eta^2} \, ,
  \qquad
  0 <  \epsal  \leq 1 \, .
\end{align}
The action is then given by
\begin{align} 
    S &= \Omega \frac{m^{6-D}}{4\eta^2} S_C(\epsal) 
= \Omega \frac{m^{6-D}}{4\eta^2}  \int_0^\infty {\rm d}\tilde\rho \  \tilde\rho^{D-1}  
  \left( \frac{1}{2} \left( \frac{{\rm d}\varphi}{{\rm d}\tilde\rho} \right)^2 + 
  \tilde V(\varphi) \right) 
\end{align}
with the dimensionless potential
\begin{equation} \label{VtC}
   \tilde V(\varphi) = \frac{1}{2} \varphi^2 + \frac{1}{2} \varphi^3 + 
   \frac{1-\epsal}{8} \varphi^4 \, ,
\end{equation}
and with $S_C(\epsal)$ a function of $\epsal$ only.

The thin-wall regime, where the minima are almost degenerate, corresponds to $\epsal\simeq 0$.
Here, one can perform an expansion in powers of $\epsal$ for the scalar field, solve the equations of motion order by order in $\epsal$ and plug the bounce solution into the Euclidean action.
The result, up to order $\epsal^2$, is given by
\begin{equation}
S_C^{(2)}(\epsal)=\left(\frac{D-1}{3\epsilon_\alpha}\right)^{D-1}\frac{2}{3D}
    \left(1+
    \epsilon_\alpha\frac{3D+8}{2}
    +\epsilon_\alpha^2\frac{9D^3-11D^2+138D-12D\pi^2-64}{8(D-1)}\right) \, .
\end{equation}
Notice that this result was obtain analytically, as explained in~\cite{Matteini:2024xvg}\footnote{Up to order $\epsal^4$, the calculation can be performed fully analytically. See also~\cite{Ivanov:2022osf} for a similar derivation. Moreover, in~\cite{Matteini:2024xvg} the authors were able to obtain a semi-analytical result up to order $\epsal^{16}$, by deriving the equations of motion at each order analytically and solving them numerically.}.
There, it was also shown that truncating the expansion at this order produces a result for the Euclidean action that agrees well with the numerical result over a large portion of the $\epsal$ parameter space, well beyond the strict thin-wall regime.
%
This serves as motivation to exploit the analytical control given by the thin-wall approximation to explore more phenomenologically relevant scenarios, such as the coupled fluid-scalar field model presented in the next section.
%

\section{Coupled fluid-scalar field model} \label{sec:FF}
\begin{figure}
  \centering
  \includegraphics[width=0.75\textwidth]{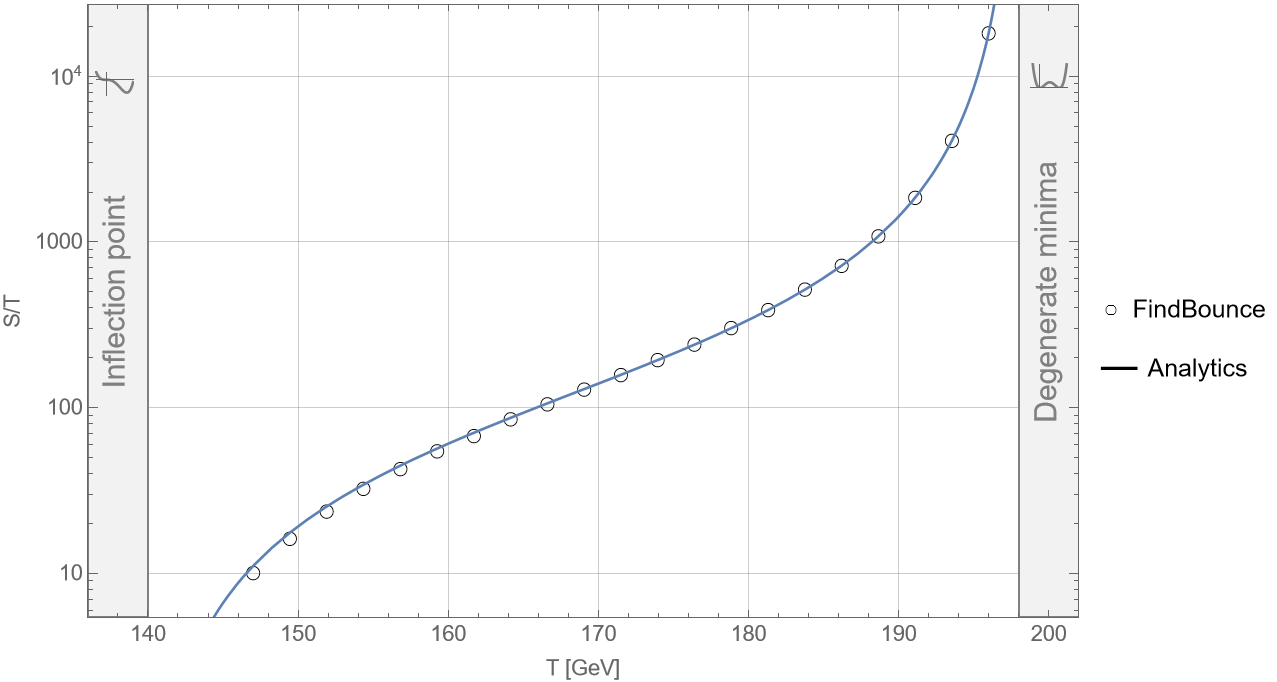}
  \caption{Comparison between the numerical action and analytical thin-wall action, for the benchmark studied in the main body. There is excellent agreement over the whole range of temperature.}
  \label{fig:ActionFF}
\end{figure}

As an application of the analytical results presented in the previous section, I consider the coupled fluid-scalar field model (see for example~\cite{Hindmarsh:2015qta,Hindmarsh:2013xza}), which describes a cosmological phase transition driven by a scalar field coupled to the primordial plasma.
The potential is given by
\begin{equation}
V(\phi,T)=\frac{1}{2}\gamma\left(T^2-T_0^2\right)\phi^2-\frac{1}{3}AT\phi^3+\frac{1}{4}\lambda\phi^4 \, , 
\end{equation}
where the temperature-dependent coefficients parametrize the interaction between the scalar and the plasma.
The phase structure of the model is straightforward to obtain.
%
The critical temperature is given by
\begin{equation}
T_c=\frac{\sqrt{9\gamma\lambda}}{\sqrt{9\gamma\lambda-2A^2}} \, T_0 \, ,
\end{equation}
while the inflection point is reached at the temperature $T_0$.
This potential can be directly mapped to the thin-wall potential (\ref{eq:VTW}) via
\begin{equation}
m^2=\gamma\left(T^2-T_0^2\right) \, , \qquad  \eta=-\frac{1}{3}AT \, , \qquad  \lambda_{C}=2\lambda \, , \qquad \epsal=1-\frac{18\lambda\gamma\left(T^2-T_0^2\right)}{4A^2T^2} \, ,
\end{equation}
so that the Euclidean action depends explicitly on temperature.
I select a specific benchmark point taken from~\cite{Hindmarsh:2015qta}:
\begin{equation}
\gamma = \frac{1}{18} \, , \qquad A=\frac{\sqrt{10}}{72} \, , \qquad \lambda=\frac{10}{648} \, , \qquad T_0=140 \text{ GeV}  \, .
\end{equation}
From Figure~\ref{fig:ActionFF}, one can see that the analytical action matches very well the numerical results, obtained using the \texttt{FindBounce} Mathematica package~\cite{Guada:2020xnz,Guada:2018jek}, over the whole range of temperatures between $T_0$ and $T_c$.
Moreover, the nucleation temperature and the strength agree with~\cite{Hindmarsh:2015qta}, as shown in Table~\ref{tab:benchmark}.

\begin{table}[ht]
\centering
\begin{tabular}{ |c||c|c|c| } 
 \hline
 Method &  $T_c$ [GeV] & $T_N$ [GeV] & $\alpha_{T_N}$ \\ 
 \hline
 Simulation & $\sqrt{2} \, T_0=197.99$ & $0.86 \, T_c=170.27$ & 0.01 \\
 Analytical & 197.99 & 170.22 & 0.010 \\ 
 \hline
\end{tabular}
\caption{The results for the "Simulation" method are taken directly from~\cite{Hindmarsh:2015qta}. The "Analytical" results are obtained by inserting the analytical action into the definition of the nucleation temperature (\ref{eq:Tn}), and by evaluating equation (\ref{eq:alpha}) for the strength at such temperature.}
\label{tab:benchmark}
\end{table}

\section{Conclusion} \label{conclusion}
In this work I have shown that the analytical results that can be obtained for a model consisting of a single scalar field using the thin-wall approximation have a wider range of validity than it is usually appreciated.
Moreover, these results can be mapped to phenomenologically relevant scenarios for cosmological phase transitions, provided the potential is given by a quartic polynomial.
As an explicit example, I considered the widely used coupled fluid-scalar field model.
I compared results from numerical simulations of cosmological phase transitions from the literature to analytical predictions given by the thin-wall approximation, and obtained excellent agreement.

\acknowledgments
I gratefully acknowledge the contribution of 
Miha Nemev\v{s}ek, Yutaro Shoji and Lorenzo Ubaldi to the original work that this proceeding is based on. MM is supported by the Slovenian Research Agency’s young researcher program under grant No. PR-11241.

\end{document}